\documentclass[namedreferences,hyperref,optionalrh]{spr-sola}
\usepackage{graphicx}        
\usepackage{color}           

\chardef\us=`\_
\begin{document}

\begin{frontmatter}


\title{Shock Signatures of the Successive Type-II Solar Radio Bursts at Meter Wavelength\\ {\it Solar Physics}}

\author[addressref={aff1},email={vasanth.veluchamy@uj.edu.pl}]{V.~\surname{Vasanth}\orcid{0000-0002-6056-7899} \sep}

       \address[id=aff1]{Astronomical Observatory of Jagiellonian University, Krakow 30-244, Poland}

\runningauthor{Vasanth}
\runningtitle{Shock Signatures of the Successive Type-II Solar Radio Bursts at Meter Wavelength}

\begin{abstract}
The successive type-II solar radio bursts observed on 31 July 2012 by the Bruny Island Radio Spectrometer (BIRS) in the frequency range between 62 $\--$ 6 MHz is reported and analyzed. The first type-II radio burst shows clear fundamental and harmonic band structures, while only one band is observed for the second type-II radio burst and is considered as the harmonic band. The first type-II radio burst is observed in the frequency range of 57 $\--$ 27 MHz between 00:03 $\--$ 00:09 UT at the harmonic band. The second type-II burst is observed between 00:18 $\--$ 00:27 UT in the frequency range of 43 $\--$ 17 MHz. The type-II radio bursts are associated with a C6 class flare located at the south$\--$eastern limb (S24E87) and a CME observed from STEREO and LASCO observations. The EUVI signatures of the CME is observed in the ST$\--$B EUVI FOV between 23:56 (on 30 July 2012) to 00:06 UT (on 31 July 2012), and are observed in the ST-B COR1 FOV between 00:10 $\--$ 00:35 UT moving within an average speed of 725 $\pm$ 101 km $s^{-1}$. The CME is observed in the LASCO C2 FOV after 00:12 UT as a partial halo CME moving with an average speed of 486 km s$^{-1}$. The height-time plot shows that the first type-II radio burst was formed by the CME-shock along the shock front and the second type-II radio burst along the shock-dip structure, probably the dip structure results from the shock transiting across the high dense streamer structure. The successive type-II bursts are most likely produced by the single CME shock and their interactions with the streamer structures. The first type-II radio burst by the CME shock and the second type-II radio burst by the CME shock$\--$streamer interactions.

\end{abstract}

\end{frontmatter}

\section{Introduction}
     \label{S-Introduction}

Type-II solar radio bursts are intense, narrow band slow drifting emission observed in the dynamic spectra from higher to lower frequencies, generated by the conversion of plasma waves excited by the electrons accelerated at the MHD shocks propagating outward from the Sun \citep{Nelson85}. The high frequency components in the decimetric to metric wavelengths are observed using ground based radio telescopes and the decameter-kilometer wavelengths from the space based instruments due to the ionospheric cut-off frequencies. The type-II radio bursts are first identified in the meter wavelength by \cite{Payne-Scott47} and later \cite{Wild50} named them slow-drifting burst. In the radio dynamic spectrum, they are predominately detected in two frequency bands that corresponds to the local plasma frequency and its harmonics \citep{Payne-Scott47, Wild50, Nelson85, Vasanth11, Vasanth14, Vasanth15, Prakash12, Shanmugaraju18}. The solar eruptions such as solar flares and CMEs are efficient in accelerating electrons and triggering shocks for the generation of the type-II radio bursts \citep{Nelson85, Classen02, Shanmugaraju03, Vrsnak08, Pohjolainen08, Vasanth13a, Vasanth13b, Chen14, Kong12, Feng13, Shanmugaraju18, Vasanth24, Vasanth25}.

  Occasionally, type-II radio bursts are observed in succession with two or more radio emission components termed as multiple type-II radio bursts, that typically occurs within 30 minutes following the first type-II radio burst. The first observations of multiple type-II radio bursts were reported by \cite{Robinson82} and their results supports that the single shock wave and their travelling disturbances through different coronal regions could produce multiple type-II radio bursts. While, \cite{Klassen99} proposed that the subsequent type-II burst could be triggered by evaporation shocks. Multiple type-II bursts are studied by several authors \citep{Shanmugaraju03, Shanmugaraju05, Subramanian06, Pohjolainen08, Cho11, Ramesh23}. Most authors supported that the multiple type-II bursts are produced by the single shock wave propagating along the boundaries of the different coronal structures. Some studies, compared the estimated shock velocities of multiple type-II radio bursts with the CME speeds and concluded that the first type-II burst is driven by shock along the CME front and subsequent type-II burst by shock along the flank region \citep{Shanmugaraju03, Subramanian06}. \cite{Karlicky94} examined the chance of existence of two separate shock waves by a single flare like disturbances, with the first one being short-lived  blast wave and the second one  a longer piston driven shock associated with chromospheric evaporation. \cite{Wagner83}, suggests that solar eruptions can generate two shock waves, one preceding the CME and another blast wave propagating through the CME.


The present study reports the successive type II radio bursts occurred on 31 July 2012 observed by the Bruny Island Radio Spectrometer (BIRS; \cite{Erickson97}) in the frequency range between 62 - 6 MHz with the time resolution of 3s and delay of 8 minutes between the first and second radio bursts. It is well known that the BIRS is designed to observed the radio bursts at low frequency below the ionospheric cut-off frequencies upto 3 MHz and is in operation until 2015. In Section 2, provides the overview of the events: observations at the EUV, white-light, X-ray and radio bands are provided, section 3 present the results by furnishing the kinematic information of radio bursts and their eruptive structures in the EUV and white light observations in the corona, section 4 for discussions and section 5 conclude their results.

\section{Overview of the Event}
    \label{S-Overview of the Event}
\subsection{Radio Observations}

   Two successive type-II solar radio bursts were observed by the Bruny Island Radio Spectrometer (BIRS: \cite{Erickson97})
   in the frequency range between 62 $\--$ 6 MHz with a time cadence of 3s on 31 July 2012 as shown in Figure1.
   Most ground based radio observatories has lower frequency observations around or until 20 $\--$ 10 MHz due to ionospheric cut-off frequencies,
   while the BIRS located at Tasmania has low ionospheric densities and free from locally generated interferences
   therefore, sometimes it has observations up to or until 3 MHz.

    \begin{figure*}[h]  
     \vspace{-0.01\textwidth}
      \centerline{\hspace{0.01\textwidth} \includegraphics[width=1.0\textwidth,angle=0]{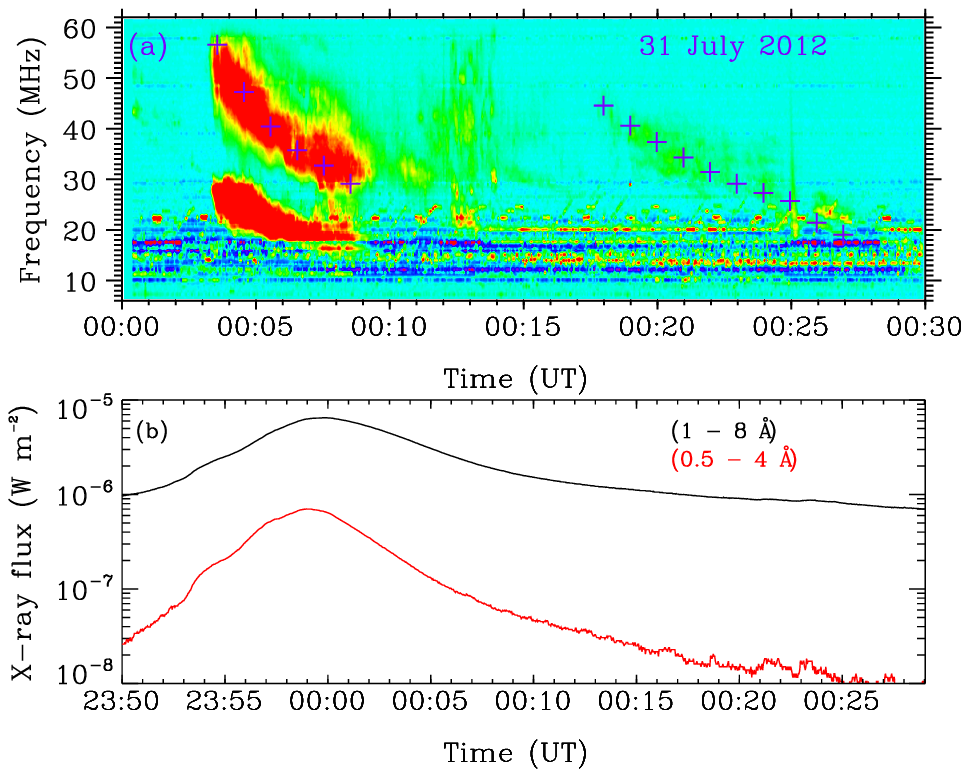}
                    }
    \vspace{0.03\textwidth}
        \caption{(a) The radio dynamic spectrum of the successive type-II radio bursts recorded by the \textrm{BIRS} (62 $\--$ 6 MHz) on 31 July 2012 between 00:00 UT $\--$ 00:30 UT, (b) \textrm{GOES} X-ray profile at 0.5 $\--$ 4 and 1 $\--$ 8 {\AA}.}
    \end{figure*}

   The characteristics of the observed successive type-II bursts are in agreement with the one shown and reported
   by the pervious studies \citep{Robinson82, Shanmugaraju05, Subramanian06, Vasanth025}, with a close time sequence
   of 8 mins between them, also with the different drift rates. The first burst has a clear fundamental and harmonic band structures,
   while the second burst has only one band and is considered as the harmonic band. The harmonic band is used for the
   analysis and several frequency points are selected on the harmonic band (see Figure 1). The first burst started at
   00:03 UT from 57 MHz with the frequency drift rate of -0.085 MHz s$^{-1}$, and the second type-II burst appeared at
   00:18 UT from 43 MHz with the frequency drift rate of -0.05 MHz s$^{-1}$. The first burst drifts faster than the
   second burst as reported by the pervious studies \citep{Robinson82, Shanmugaraju05, Cho11}.

   The detailed structure of the burst shows that the first type-II burst appears to be more intense and broader
   compared to the second type-II burst, while the duration of second type-II burst is larger than the first type-II
   burst. The different drifting nature of the successive type-II bursts suggest that they are  accelerated at
   different coronal structures or regions. The first type-II burst appears in the frequency range of 57 $\--$ 27 MHz
   between 00:03 $\--$ 00:09 UT, and the second type-II burst appears in the frequency range of 43 $\--$ 17 MHz between
   00:18 $\--$ 00:27 UT. It is also important to note that there is a group of type-III radio bursts observed after first
   type-II burst around 00:12 UT $\--$ 00:16 UT, the second type-II burst appears to following the group of type-IIIs. The
   type-III radio bursts are produced by the electron beam propagating along the open magnetic field line from the Sun to the outer corona.

\begin{figure*}    
   \centerline{\includegraphics[width=1.0\textwidth,clip=]{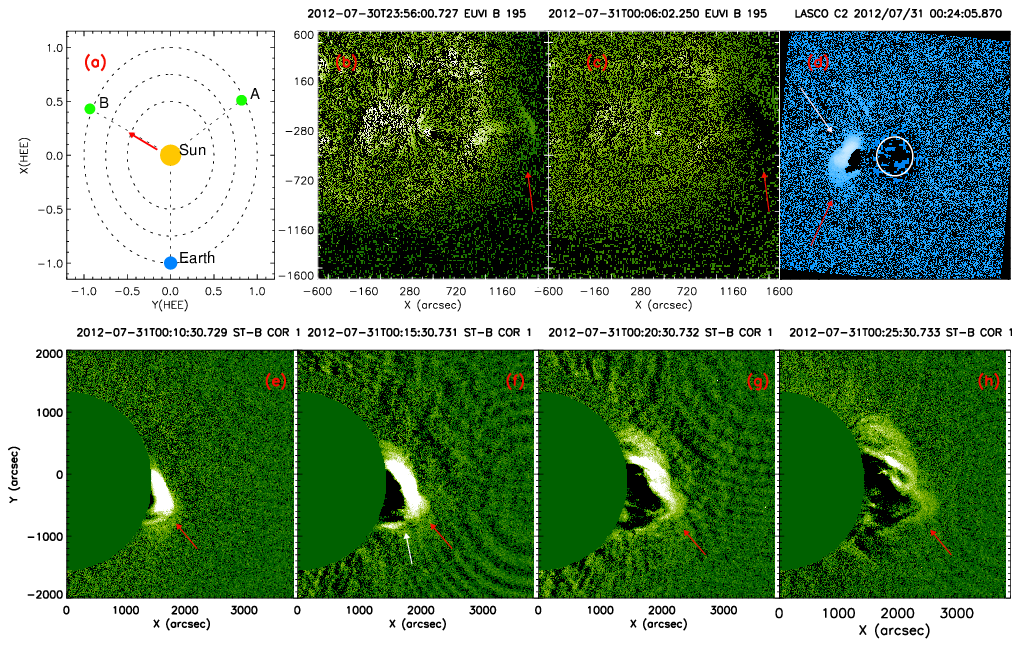}
                }
   \caption{(a) Relative location of the twin spacecraft (\textrm{STEREO}) and the Earth, (b) - (c) sequence of the ST-B EUVI running difference images showing the eruptive structure at 195 {\AA}, (d) the CME and streamer structure observed in the LASCO C2 difference image, (e) - (h) the temporal and structural evolution of CME observed by the ST-B COR1 running difference images. The red arrow at (b) - (h) indicates the CME structure, the white arrow at (d) shows the brightness enhancement observed partially along the CME front and at (f) indicates the dip structure.}
  \end{figure*}

   \subsection{EUV-White light eruptive structures}

   The successive type-II radio bursts are closely associated with a CME observed in the twin Solar Terestrial Relations Observatory (STEREO: \cite{Kaiser08}) and the Large Angle Spectrometric Coronagraph (LASCO: \cite{Brueckner95}) C2 on board the Solar and Heliospheric Observatory (SOHO: \cite{Domingo95}) observations. The CME starts to erupt around 23:56 UT first observed in the ST-B EUVI FOV during the impulsive phase of a C6-class flare (see Figure1b) that started at 23:50 UT, peaked  at 23:59 UT and ends at 00:04 UT located at the south-eastern limb (S24E87) from the active region AR11538. During a failed solar eruption, streams of plasma or filament material initially moves upward rising to a certain height and then fall back to the Sun. They don't have sufficient energy to exceed the threshold to be erupted. So they don't have a CME. Based on the multiwavelength observations and their magnetic field configurations of the eruptive structures, there are no signatures of multiple eruptions either flares and$/$or CMEs associated with the successive type-II radio bursts. There is no evidence on the movement of filament materials or plasma streams eviction from the active regions site, probably there is no clear sign of failed eruptions associated with the present event.

   The CME was simultaneously detected in STEREO and SOHO FOV at 00:24 and 00:25 UT, which allow us to identify it 3-D position, there is a streamer structures located and major part of the CME pass through the streamer structures clearly seen in the LASCO FOV. The CME-streamer interactions can be seen after 00:10:30 UT in the ST-B COR1 images (see Figure 2). The sequences of the CME-streamer structures are shown in Figure 2 between 00:05 $\--$ 00:25 UT. The first contact of the CME-streamer might have occurred after 00:10:30 UT, after that the type-III radio burst and the second type-II radio burst are observed (see Figure 1). The PFSS magnetic field lines are shown over the LASCO-EIT and the LASCO-C2 images in Figure 3. The streamer regions that is the high located closed loop structure together with the open field structure appears clearly in the LASCO-C2 images. The major part of the CME passes through the streamer structure. The CME first appears in the LASCO-C2 FOV around 00:12 UT, while the CME-streamer interactions observed in the ST-B COR1 image around 00:10:30 UT. The density in the corona is also measured using the ST-B COR1 $P_B$ data and the estimated density around 00:25 UT decrease from $10^{7}cm^{-3}$ as shown in Figure 4, agrees well with the observed frequency of the successive type-II radio bursts.

   \begin{figure*}   
         \centerline{\includegraphics[width=0.95\textwidth,clip=]{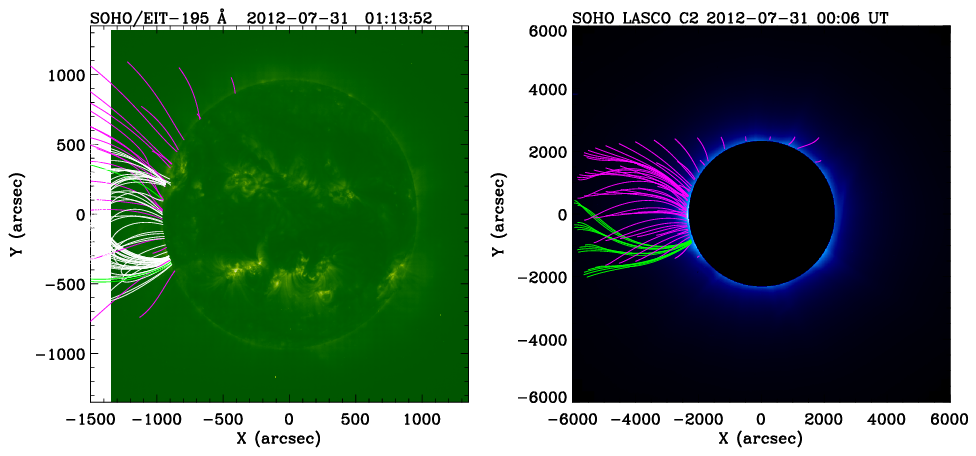}
                    }
        \vspace{-0.005\textwidth}
        \caption{(a) The PFSS magnetic field lines shown over the (a) LASCO-EIT image and (b) LASCO C2 image. The major part of the CME passes through the streamer structure.}
    \end{figure*}

     \begin{figure*}[h]   
      \centerline{\includegraphics[width=1.1\textwidth,clip=]{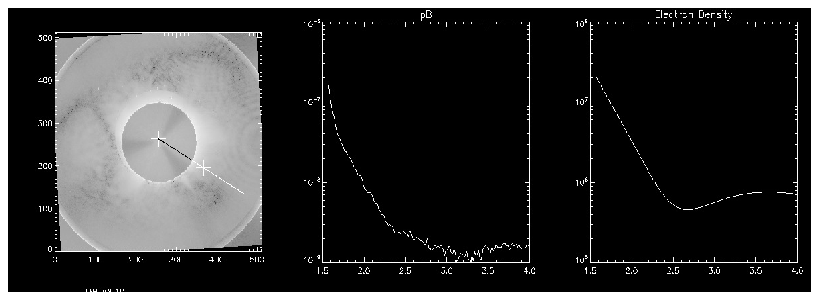}
                    }
    \vspace{-0.05\textwidth}
    \caption{Temporal evolution of the electron density profile as a function of the distance obtained from the ST-B COR1 $P_B$ data at 00:25 UT.}
   \end{figure*}

   The 3-D reconstruction of the eruptive structures are obtained from the GCS method in pythea (see Figure 5), they also provides the heights of the Shock and flux rope from which one can determine the shock stand-off distance ($\Delta$R = $R_S$ - $R_f$), the radius of curvature of the flux rope ($R_c$) is obtained by fitting a circle. Using the $\Delta$R and Rc, one can get the Alfv{\'e}n mach number ($M_A$). For $R_S$ = 3.39, $R_f$ = 3.22, $R_c$ = 0.644, $\Delta$R/$R_c$ = 0.263, gives $M_A$ = 2.533 for $\gamma$ = 4/3 and $M_A$ = 3.717 for $\gamma$ = 5/3 (where $\gamma$ - specific heat ratio (polytropic index)). The measurements are listed in Table 1.

   It is to be noted that there are two COR1 observations available between 00:05 $\--$ 00:10 UT, providing a speed of 1253 $\pm$ 313 km $s^{-1}$  for the first type-II radio burst and three COR1 observations are available between 00:20 $\--$ 00:30 UT, providing a speed of 835 $\pm$ 139 km $s^{-1}$ for the second type-II radio burst. We used the mean Alfv{\'e}n mach number ($M_A$) corresponding to $\gamma$ = 4/3 from the above measurements (see Table 1) to estimated the Alfv{\'e}n speed ($V_A$) and the magnetic field strength (B). Table 1 provides the information of shock and flux rope measurements obtained using Pythea. In Table 1, column 1 denotes the time in UT, column 2 indicates the height of the shock ($R_S$), column 3 represents the height of flux rope ($R_f$), column 4 specifies the shock stand-off distance ($\Delta$R = $R_S$ - $R_f$), column 5 shows the radius of curvature of the flux rope ($R_c$), column 6 denotes the relative stand-off distance ($\delta$ = $\Delta$R/$R_c$), column 7 shows the Alfv{\'e}n mach number ($M_A$) for $\gamma$ = 4/3 and column 8 shows the Alfv{\'e}n mach number ($M_A$) for $\gamma$ = 5/3.

    \begin{figure*}[h]   
       \centerline{\includegraphics[width=0.9\textwidth,clip=]{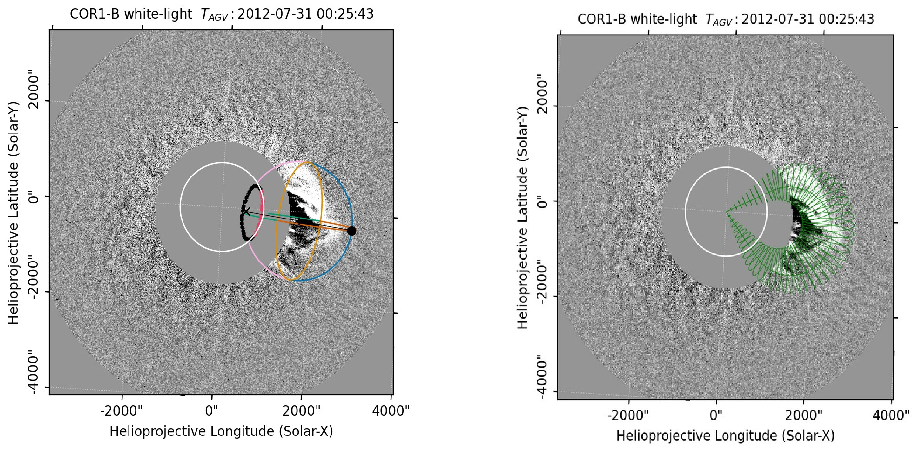}
       }
    \caption{The heights of shock and flux rope measured using the Pythea (GCS and Ellipsoidal models)}.
   \end{figure*}

   \begin{equation}
   \Delta R/R_c = 0.81[(\gamma - 1)M^2]/[(\gamma + 1)(M^2 - 1)]
\end{equation}

\begin{equation}
 M^2 = 1 + [1.24 \delta - (\gamma -1)/(\gamma + 1)^{-1}
\end{equation}

 Once $M_A$ is known, Alfv{\'e}n speed ($V_A$) can be estimated by using the relation $V_A$ = $V_S$/$M_A$, then the magnetic field (B) can be derived using

\begin{equation}
 B = 5.1 * 10^{-5} V_A f (G)
\end{equation}

\begin{table}
\caption{ The shock and flux rope measurements obtained using Pythea.
}
\label{T-simple}
\begin{tabular}{lllllllll}     
\hline
 \multicolumn{9}{|c|}{\textbf{First type-II radio burst}} \\
  \hline                   
\textbf{Time (UT)} & \textbf{$R_S$} & \textbf{$R_f$} & \textbf{$\Delta$R} & \textbf{$R_c$} & \textbf{$\Delta$R/$R_c$} & \textbf{$M_A$ ( $\gamma$ = 4/3)} & \textbf{$M_A$ ( $\gamma$ = 5/3)} & \\
\hline
00:05 & 1.99 & 1.91 & 0.08 & 0.235 & 0.34 & 2.137 & 2.557 & \\
00:10 & 2.53 & 2.30  & 0.23 & 0.383 & 0.601 & 1.629 & 1.735 & \\
\hline
Mean &       &       &      &        &       & 1.883  & 2.146 & \\
\hline
\multicolumn{9}{|c|}{\textbf{Second type-II radio burst}} \\
 \hline
00:20 & 2.97 & 2.74  & 0.23 & 0.548 & 0.419 & 1.909 & 2.164 & \\
00:25 & 3.39 & 3.22  & 0.17 & 0.644  & 0.263 & 2.533 & 3.717 & \\
00:30 & 3.69 & 3.50  & 0.19 & 0.7  & 0.271 & 2.478 & 3.517 & \\
\hline
Mean &       &       &      &        &       & 2.306  & 3.132  & \\
  \hline
\end{tabular}
\end{table}

\section{Results}

   \vspace{0.05\textwidth}
\begin{figure*}[h]   
         \vspace{0.05\textwidth}
        \centerline {\hspace*{-0.05\textwidth} \includegraphics[width=0.75\textwidth]{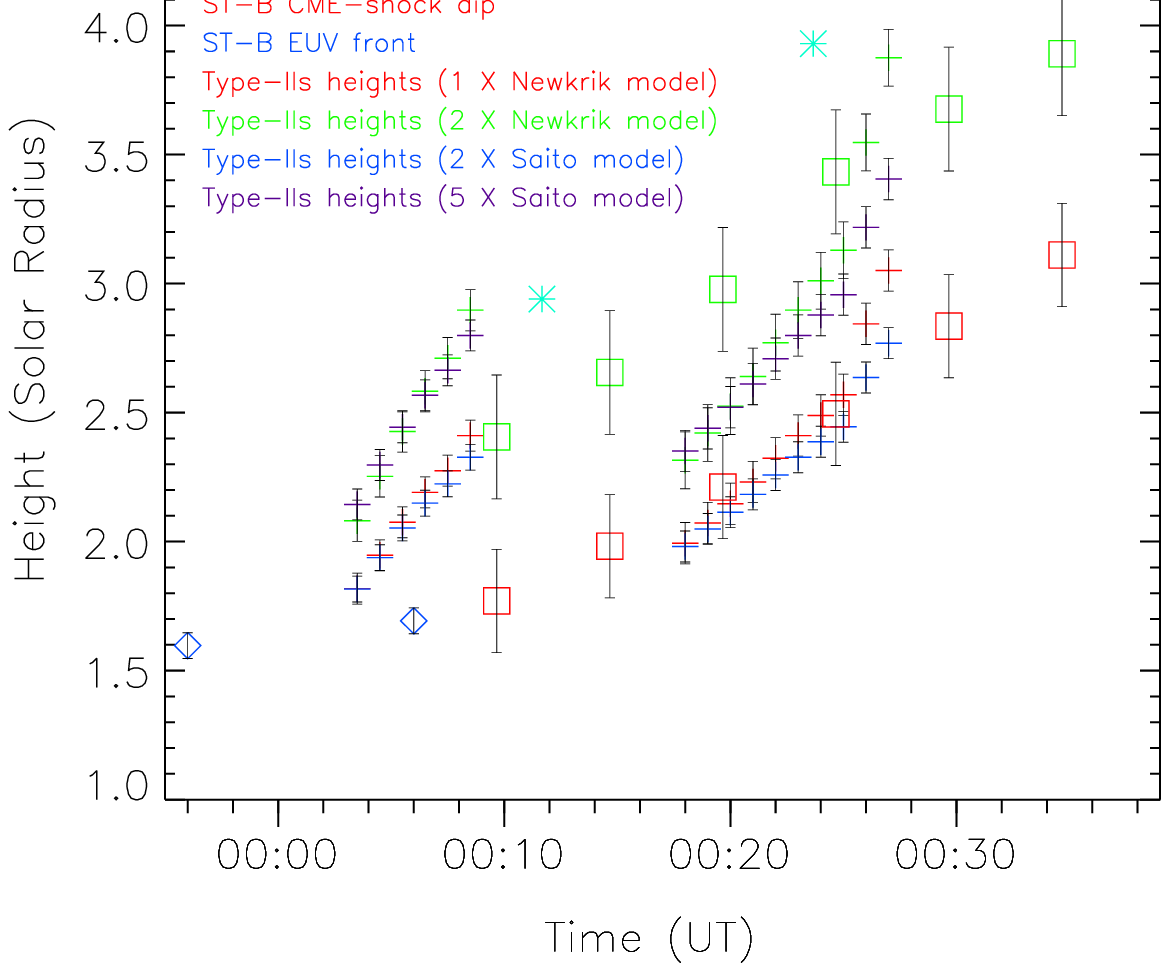}
        }
    \vspace{0.05\textwidth}
    \caption{The height-time profiles observed at EUV, white-light and radio wavelengths.}
    \end{figure*}

   The relationship between CME and successive type-II radio bursts are explored by comparing the heights of the different structures.

   The type-II radio bursts observed during the CME eruptions are signatures of coronal shock waves,
   the radio emitting frequencies could be converted into heights by applying the appropriate coronal
   electron density models. Then by comparing the type-II height obtained from the frequency
   drifts with the CME heights, the driver of the shock can be determined. To do so, we selected several
   frequency points on the harmonic bands of the first and the second type-II radio burst (See Figure 1), then
   by applying 1x and 2xNewkirk density model \citep{Newkirk61}, as well as 2x,and 5xSaito density model \citep{Saito70}, the heights are
   estimated and shown in Figure 6. The height-time plots shows the heights at various wavelengths,
   EUV, white light and radio observations. The heights of the first bursts matches well with the
   CME-frontal structure and the height of the second type-II burst appears along the CME shock-dip
   structure. Probably, the dip structure is the outcome of shock transiting across the high dense
   low level loop structure i.e. along the streamer structure. The high dense streamer region acts as a good source region with low
   Alfv{\'e}nic region for the generation of the coronal shock. Therefore the second radio burst should be an outcome of the CME shock-streamer interactions.
   The CME driven shock encountering a streamer structure results in local steepening, that increases the compression ratio and form the brightness enhancement region. These structures forms an efficient trap for the repetitive particle acceleration. We observe a dip along the flank structure of ST-B COR1 CME and the dip indicates the shock has transited across the streamer structure with a strong compression or interaction that leads to the generation of efficient electron acceleration and produced the second type-II radio burst. There is also brightness enhancement observed partially along the CME-front in LASCO C2 FOV indicating the strong compression caused by the propagation of shock into the dense loops.

  The precise shock location on the eruptive structures or CMEs can be provided using the radio imaging observations from radio heliograph data. Due to lack of radio imaging observations, we cannot provide precise location. But can suggest the possible source of shocks on the eruptive structures. The first type-II radio burst might be produced along the CME-front and the second type-II burst along the shock-dip on the CME-flank that interacts with the high dense streamer structure.
    The estimated shock speed of the first type-II radio burst for 1xN (1376 $\pm$ 107 km $s^{-1}$), 2xN (1893 $\pm$ 116 km $s^{-1}$)
   in the height range of 1.82 $-$ 2.9 $R_\odot$ , and for 2xS (1186 $\pm$ 95 km $s^{-1}$), 5xS (1520 $\pm$ 114 km $s^{-1}$)
   in the height range of 1.82 $-$ 2.8  $R_\odot$, the estimated speed of first type-II radio burst using 1xN and 2xS is in agreement with the CME speed (1253 $\pm$ 313 km $s^{-1}$) obtained using GCS method, the estimated Alfv{\'e}n speed of the first type-II radio burst for
   1xN (732 $\pm$ 57 km $s^{-1}$), 2xN (1007 $\pm$ 62 km $s^{-1}$) and for 2xS (631 $\pm$ 50 km $s^{-1}$), 5xS (808 $\pm$ 61 km $s^{-1}$),
   their magnetic field strength (B) for 1xN (0.69 $\pm$ 0.09 G), 2xN (0.95 $\pm$ 0.1 G), 2xS (0.60 $\pm$ 0.09 G) and 5xS (0.77 $\pm$ 0.11 G).
   Similarly the estimated shock speed of the second type-II radio burst for 1xN (1362 $\pm$ 279 km $s^{-1}$), 2xN (2011 $\pm$ 447 km $s^{-1}$)
   in the height range of 1.99 $-$ 3.88 $R_\odot$ , and for 2xS (1016 $\pm$ 174 km $s^{-1}$), 5xS (1358 $\pm$ 246 km $s^{-1}$)
   in the height range of 1.98 $-$ 3.41  $R_\odot$, while the CME speed obtained from GCS method is 835 $\pm$ 139 km $s^{-1}$,
   the estimated Alfv{\'e}n speed of the second type-II radio burst for
   1xN (592 $\pm$ 121 km $s^{-1}$), 2xN (874 $\pm$ 253 km $s^{-1}$) and for 2xS (442 $\pm$ 75 km $s^{-1}$), 5xS (591 $\pm$ 107 km $s^{-1}$),
   their magnetic field strength (B) for 1xN (0.4 $\pm$ 0.05 G), 2xN (0.59 $\pm$ 0.1 G), 2xS (0.31 $\pm$ 0.03 G) and 5xS (0.41 $\pm$ 0.04 G).
   The values of 1xN and 2xS as well as 2xN and 5xS agrees well to each other i.e. lies closer to each other.

\section{Discussion}

   The present study supports the scenario of single shock wave and their travelling disturbances along the different regions of the coronal structures as the cause for the generation of the successive type-II radio bursts. These results are supported by the h-t plot, that the first type-II burst are formed by the CME-shock along the shock front and the second type-II radio burst along the shock-dip structure. The dip structure formed by the shock transit across the high dense streamer structure. This indicates that the CME shock-streamer interaction might be responsible for the generation of the second type-II burst. The successive type-II radio bursts are two or more type-II burst components appearing in sequence over a closest time in the radio dynamic spectrum, consisting of different drift rates. The different drift rates indicates different propagating speeds that originated from different source regions and sometimes one of their speed can matches the CME speed. The type-II solar radio bursts observed during the eruptions (flares or CMEs) indicates the formation of shock waves. Therefore for the case of multiple or successive type-II radio bursts one expects multiple shock waves. Majority of the earlier observational results supports that the successive type-II radio bursts arise from single shock wave and their propagating disturbances encountering different regions in the corona with distinct plasma and magnetic properties \citep{Robinson82, Shanmugaraju03, Shanmugaraju05, Subramanian06, Cho11}. In certain cases, If a CME associated eruptive flare undergoes consecutive impulsive phases, then each phase has the potential to drive a radio emission, thereby acting as a source for generating the successive type-II radio bursts. Most of the previous studies based on the statistical analysis supported the single shock wave as a source for successive type-II bursts, further examining the correlation between CME speed and type-II speed, it become evident that the radio emitting energetic electrons are concentrated along the front and flank of the CME. In the present case, the first type-II burst is generated along the CME shock front and the second type-II burst along the dip region resulting from CME shock-streamer interaction.


 The possibility for the existence of two kinds of coronal shock waves were first proposed by \cite{Maxwell82} from observations and \cite{Gergely84} from theoretical assumptions. Follow it, there are lot of theoretical or simulation studies are carried out with the assumption of two shocks \citep{Karlicky94, Magara00, Sakai05}. Few observational studies also supports two coronal shock waves: \cite{Wagner83}, one preceding the CME and another blast wave propagating through the CME, \cite{Mancuso04} concluded that the shock waves can be originated at discrete locations along front and flank of a CMEs. It is important to note that none of the above studies has two CME eruptions or multiple CME shocks.

  Recently, \cite{Vasanth025} reported for the first time, the direct observations of two coronal shocks from two CMEs generating the successive type-II radio bursts. The two CMEs originated from the same active region produced the two shock signatures, further confirmed by the existence of two type-II radio bursts. According to  \cite{Robinson85}, consecutive flaring events are more likely to generate successive type-II radio bursts, with one near the CME-leading edge and another during the impulsive phase of the flare. If an eruption is associated with flares and CMEs, then both flares and CMEs can contribute to the formation of shock waves. Therefore, the occurrence of two independent coronal shocks from two different drivers are strongly influenced by the properties of the surrounding plasmas and their local conditions. More studies on the observational aspects of multiple or successive type-II radio bursts are required to verify the multiple CMEs shocks accelerations and their physical condition triggering it. It might help us to achieve a better understanding of the successive type-II radio bursts and their shock acceleration mechanism.

\section{Conclusions}

  A detailed case study of the successive type-II solar radio bursts observed on 31 July 2012 by the BIRS in the frequency range of 62 $\--$ 6  MHz, with a time resolution of 3s is reported. The successive type-II radio bursts are accompanied by a C6 class flare and a partial halo CME from the south-eastern limb.

  The first type-II radio burst appears in the frequency range of 57 $\--$ 28 MHz between 00:03 $\--$ 00:09 UT and the second type-II radio burst is observed in the frequency range of 43 $\--$ 17 MHz between 00:18 $\--$ 00:27 UT. The first type-II burst appears to be more intense and broader compared to the second type-II burst and the second type-II burst exists for longer duration compared to the first type-II burst. The first type-II burst drifts faster compared to the second type-II burst. The different drift rates of the radio bursts indicates that they exists from the different source regions.

  The CME associated with the successive type-II bursts appears to be a partial halo CME clearly observed in the ST and the LASCO FOV. The major part of the CME passes along the streamer region well observed by the LASCO C2 FOV. The brightness enhancement appearing partially along the CME-front indicates the strong compression caused by the propagation of shock into the high dense loops. The existence of streamer region is shown by the PFSS magnetic field configuration on the LASCO-EIT and the LASCO-C2 images on Figure 3. The CME-streamer interaction appears on or after 00:10:30 UT on the ST-B COR1 images. We observe a dip region along the CME-flank on ST-B COR1 FOV and the dip indicates the shock has transited across the streamer structure with a strong compression or interaction that leads to the generation of efficient electron acceleration and produced the second type-II radio burst.

  The estimated speed of the first and the second type-II radio bursts are about 1100 $\--$ 2000 km $s^{-1}$ for the different multiplier of the employed Newkirk and Saito electron density models and is two or more times larger than the estimated CME speed in the ST-B COR1 FOV (725 $\pm$ 101 km $s^{-1}$). Only one CME is observed during the entire period, The h-t plot shows that the first type-II burst was formed by the CME-shock along the shock front and the second type-II radio burst along the shock-dip structure. This indicates that the CME shock-streamer interaction might be responsible for the generation of the second type-II burst, so the successive type-II radio bursts are mostly likely produced by a single CME shock and their interaction with the streamer regions. The first type-II radio burst by the CME shock and the second type-II radio burst by the CME shock-streamer interactions.

\section*{Data Availiability}
    The datasets generated and/or analyzed during the current study are available online.
    The  Bruny Island Radio Spectrometer (BIRS) data \hfill \break
 $https://www.astro.umd.edu/~white/gb/Html/Events/TypeII/20120730_235500_BIRS.html$
 GEOS soft X-ray data \hfill \break
 $https://www.ncei.noaa.gov/data/goes-space-environment-monitor/access/full/2012/07/goes15/csv/$
 STEREO-A EUVI data \hfill \break
 $https://stereo-ssc.nascom.nasa.gov/data/ins data/secchi/L0/b/img/euvi/20120731/$
 STEREO-A COR1 data \hfill \break
 $https://stereo-ssc.nascom.nasa.gov/data/ins data/secchi/L0/b/seq/cor1/20120731/$

\begin{acks}
This work was supported by the POB Anthropocene research program of Jagiellonian University, Krakow, Poland.
The author greatly acknowledge the various online data centers of
NOAA and NASA for providing the data. We express our thanks to the
Bruny Island Radio Spectrometer (BIRS), STEREO/SECCHI, SOHO/LASCO teams
for providing the data and LASCO CME catalog used is generated
and maintained by the Center for Solar Physics and Space Weather,
The Catholic University of America in cooperation with the Naval Research Laboratory and NASA.
\end{acks}

\bibliographystyle{spr-mp-sola}

\bibliography{Vasanth}

\end{document}